**Giant thermal spin torque assisted magnetic tunnel junction switching**

Aakash Pushp[1*§], Timothy Phung[1,2*], Charles Rettner[1], Brian P. Hughes[1], See-Hun Yang[1] and Stuart S.P. Parkin[1§]

[1]IBM Almaden Research Center, San Jose, California 95120, USA

[2]Department of Electrical Engineering, Stanford University, Stanford, California 94305, USA

[*]These authors contributed equally to the work.

[§]Correspondence to be addressed to: apushp@us.ibm.com (A.P.);

Stuart.Parkin@us.ibm.com (S.S.P.P.)

**Spin-polarized charge-currents induce magnetic tunnel junction (MTJ) switching by virtue of spin-transfer-torque (STT). Recently, by taking advantage of the spin-dependent thermoelectric properties of magnetic materials, novel means of generating spin-currents from temperature gradients, and their associated thermal-spin-torques (TSTs) have been proposed, but so far these TSTs have not been large enough to influence MTJ switching. Here we demonstrate significant TSTs in MTJs by generating large temperature gradients across ultrathin MgO tunnel barriers that considerably affect the switching fields of the MTJ. We attribute the origin of the TST to an asymmetry of the tunneling conductance across the zero-bias voltage of the MTJ. Remarkably, we estimate through magneto-Seebeck voltage measurements that the charge-currents that would be generated due to the temperature gradient would give rise to STT that is a thousand times too small to account for the changes in switching fields that we observe.**

Using heat to create potential gradients and charge-currents has been a very active area of research in thermoelectrics (*1*). Spin caloritronics (*2, 3*) adds a new dimension to this concept by considering the use of heat to create spin-dependent chemical potential gradients in ferromagnetic materials (*4*). Traditionally, electric current driven spin-currents have been used to transport spin angular momentum to change the magnetization of a magnetic material – a phenomenon known as spin-transfer-torque (STT) (*5-7*). Heat currents can also create spin-currents in magnetic materials; the transfer of spin angular momentum through this process has been named thermal-spin-torque (TST) (*8, 9*). A panoply of recent experiments that employ spin currents generated by heat have been reported that includes the spin-Seebeck effect observed in ferromagnetic metals (*10, 11*), semiconductors (*12*) and insulators (*13*), thermal spin injection from a ferromagnet into a non-magnetic metal (*14*), the magneto-Seebeck effect observed in magnetic tunnel junctions (*15-17*), Seebeck spin tunneling in ferromagnet-oxide-silicon tunnel junctions (*18*) and several others (*19, 20*). On the other hand, whilst there have been several theoretical predictions (*8, 9, 21, 22*) of the TST, there have been few experiments to date. In one experiment, evidence of TST was established in Co-Cu-Co spin-valve nanowires (*23*). However, in this work the same current was used for both heating and probing the device thus making it difficult to unravel the individual contribution of TST from the simultaneously generated STT.

In our device, the heating current is distinct from the probing current, which helps to decouple pure temperature gradient effects from charge current driven STT effects. We find that a temperature gradient of ~ 1 K/nm across a 0.9 nm thick MgO tunnel barrier in an MTJ induces modest charge currents of the order of $1 \times 10^3$ A/cm$^2$ along with large spin currents that induce significant TST. The TST is as large as the STT that would be created by a pure charge current density of $1 \times 10^6$ A/cm$^2$ in these devices as well as previously reported similar devices (*24*). Furthermore, the TST is strongly dependent on the orientation of the free layer with respect to the reference layer. We show that the TST can be attributed to an asymmetry in the tunneling conductance across zero bias, which is consistent with the spin accumulation in the free layer of the MTJ due to the temperature gradient across the tunnel barrier.

Figure 1 illustrates our experimental setup that consists of an MgO based MTJ(*25, 26*), grown by magnetron sputtering on thermally conducting MgO (100) substrate with the layer sequence from bottom to top as: 12.5 IrMn | 2 CoFe | 0.8 Ru | 1.8 CoFe | 0.9 MgO | 1 CoFe | 3 NiFe | 5 Ru, where the numbers represent film thicknesses in nanometers (see Supplementary Information SI for details about device fabrication). The free layer is 200 nm wide and 500 nm long whereas the reference layer of the MTJ is of considerably larger proportions (3 μm wide and 11 μm long) and serves as an on-site thermometer to measure the local temperature of the MTJ upon heating. A 1 μm wide resistive heater (resistivity ~ 2 mΩ-cm) made of ScN is deposited above the MTJ and is electrically isolated from the top contact of the MTJ by a 20 nm thick alumina (AlO$_x$) pad. The advantage of this geometry is that pure thermal gradient effects on the MTJ switching can be studied in the closed-circuit configuration (Fig. 1b) with minimal sensing current through the MTJ, $I_{MTJ}$ (= 10 µA; current density ~ $10^4$ A/cm$^2$), thereby ruling out any conventional STT or self-heating effects through the MTJ.

In order to create sharp temperature gradients for small heat input, our experiment is performed at a base temperature of 10 K that has several added advantages. Firstly, the heat capacity of the entire device is 2-3 orders of magnitude smaller than at room temperature, i.e., less amount of heat is required to raise the temperature. Secondly, the thermal conductivity (see Supplementary Information SII) of oxides is a few orders of magnitude lower at low temperatures leading to larger temperature gradients across the tunnel barrier for a given heat current. Thirdly, the resistivity of the semiconducting ScN heater is higher at lower temperatures, thereby requiring smaller heater current, $I_H$, to generate large amount of heat. Finally and most importantly, the highest temperatures of the free layer of the MTJ (< 60 K) accessed in our experiments change the saturation magnetization of the free layer by less than 1% than its lowest temperature value.

The change in the resistance of the MTJ (device I) as the magnetic field is applied to switch the free layer - the tunneling magneto-resistance (TMR) - is plotted in Fig. 2a (inset). The resistance of the MTJ when the free layer is parallel to the reference layer - the parallel (P) configuration - is lower than the resistance when the free layer is anti-parallel to the reference layer (AP configuration). Characteristic switching fields ($H_{sw}$) required to switch the free layer from P to AP configuration ($H_{sw}^{P \to AP}$) and vice versa ($H_{sw}^{AP \to P}$) are indicated. The TMR measurement can now be performed at 10 K while locally heating the MTJ with a current through the heater, $I_H$, which creates sharp temperature gradients on the order of 1 K/nm (see Supplementary Information II, III) across the tunnel barrier (transverse) and 0.1 K/nm along the length of the reference layer (longitudinal) as shown in the finite element model (Fig. 1c-e). We notice that $H_{sw}^{P \to AP}$ and $H_{sw}^{AP \to P}$ systematically change as a function of $I_H$ (Fig. 2a,b). Owing to the choice of a highly resistive heater, minimal current densities ($< 1 \times 10^6$ A/cm$^2$) are required through the heater to create such sharp temperature gradients, which amounts to small stray magnetic fields ($< 2$ Oe) at the free layer of the junction (see Supplementary Information IV). This is corrected from the $H_{sw}$ by flowing $I_H$ in both directions.

Creation of sharp temperature gradients invariably increases the absolute temperature of the MTJ. To minimize the net increase of the absolute temperature, we have grown the MTJ stack on a thermally conducting substrate, i.e., MgO (100), which acts as a heat sink for the bottom electrode of the MTJ. $H_{sw}$ also depends on the absolute temperature, $T$, of the MTJ as shown in Fig. 2e,f. Hence, in order to ascertain the TST contributions to MTJ switching, it is essential to compare the switching fields, $H_{sw}(T + \Delta T)$, measured with the temperature gradient generated from local heating with the switching fields, $H_{sw}(T)$, measured at globally elevated temperatures, where no temperature gradients ($I_H = 0$; $\Delta T = 0$) are applied. We use local thermometry (Fig. 2c,d) from the reference layer resistance, $R_{RL}$, measured as a function of $I_H$, $R_{RL}(I_H)$, and also as a function of temperature $R_{RL}(T)$ to estimate the local temperature of the MTJ with ~5 K accuracy. First, we measure the resistance change of the reference layer of the MTJ as a function of $I_H$, i.e., $\Delta R_{RL}(I_H) = R_{RL}(I_H) - R_{RL}(I_H = 0)$ at the base temperature of 10 K (Fig. 2c). Secondly, we calibrate our on device 'thermometer' by measuring $\Delta R_{RL}(T) = R_{RL}(T) - R_{RL}(T = 10K)$, as a function of temperature, $T$, and then invert the function to obtain $T(\Delta R_{RL})$, as shown in Fig. 2d. As the heater, which is 1 μm wide, heats only a section of the reference layer of the MTJ, which is 11 μm wide (see Fig. 1a,c), we can estimate the temperature of the MTJ by first scaling $\Delta R_{RL}(I_H)$ by an appropriate scale factor (see Supplementary Information SIII) before looking up the temperature for the scaled $\Delta R_{RL}$ from Fig. 2d. It may be noted that even though the temperature gradients that are created across the tunnel barrier are very large (~1 K/nm) the absolute temperature difference across the magnetic part of the stack is less than 5 K.

Comparing (Fig. 2e,f) the $H_{sw}(T)$ at elevated temperatures with the corresponding $H_{sw}(T + \Delta T)$ measured at the base temperature (10 K) with different $I_H$ scaled to the appropriate temperatures shows the evidence of the TST influencing AP → P switching, whereas no effect is seen for P → AP switching. We thus conclude that the TST originates from the vertical temperature gradient induced spin currents across the ultra-thin MgO tunnel barrier. The TST, in fact, increases $H_{sw}^{AP \to P}$ implying that the minority spins of the reference layer are accumulating into the free layer due to the temperature difference across the barrier (free layer is hotter than the reference layer), which favors the AP configuration. Such large spin accumulations are consistent with those previously observed in silicon when temperature gradients are applied across a tunnel spin injector (*18*) although we estimate that the spin currents generated in our studies are at least 10,000 times larger due to the steeper (x10) temperature gradients and significantly less resistive (x$10^6$) ultra-thin tunnel barriers in our experiments. We note that the magnitude of the spin current that depends on the flow of charge currents, will depend sensitively on the thickness of the tunnel barrier, decreasing rapidly with small increases in the thickness of the MgO tunnel barrier (*9*), perhaps explaining why no influence of temperature gradients on magnetic switching has previously been observed (*15-17*).

In order to investigate the angular dependence of the TST, similar measurements were performed on another device (device II), where the free layer was patterned at 120° to the reference layer as shown in Fig. 3a. Due to the shape anisotropy of the free layer, the TMR curve shows multiple steps corresponding to the relative orientation of the free layer with respect to the reference layer (Fig. 3b). Consequently, we can compare the $H_{sw}(T)$ and $H_{sw}(T+\Delta T)$ for the various different orientations of the free layer relative to the reference layer, and evaluate the angular dependence of the TST as shown in Fig 3c. We find that except for the AP to P switching ($H_{sw}^A$, $H_{sw}^B$ and $H_{sw}^G$ in Fig 3b), all the other $H_{sw}(T)$ and $H_{sw}(T+\Delta T)$ lie on top of each other, once again showing evidence for temperature gradient driven pure spin currents influencing MTJ switching, and also confirming that our temperature estimate of the MTJ is accurate. We note that the $H_{sw}$ for both the devices I and II reported here behave similarly – showing contribution of TST to AP → P switching but no contribution to P → AP switching.

We performed magneto-Seebeck (*15-17*) measurements (Fig. 4) on our devices to estimate the magnitude of the STT that would be obtained from thermoelectric charge currents. The maximum magnitude of the magneto-Seebeck voltage that develops across the MTJ in the open-circuit configuration (i.e. when $I_{MTJ}=0$), is for the maximum temperature gradient (maximum $I_H$) in these devices and is ~60 µV (Fig. 4b,c). This voltage would induce at most a current density, $j \sim 1 \times 10^3$ A/cm$^2$ (the Resistance-Area product of the tunnel junction is ~ 6 Ω-µm$^2$) across the tunnel barrier, which is too small to account for changes in the AP → P switching fields $\left|H_{sw}^{AP \to P}(T+\Delta T) - H_{sw}^{AP \to P}(T)\right|$ of ~ 5-10 Oe reported in our measurements. These devices require STT created by a charge current density of ~ $1 \times 10^6$ A/cm$^2$ across the tunnel barrier to change their switching fields by 10 Oe (see supplementary information SV), also consistent with a recent study (*24*) in a similar geometry. We note that self-heating from the tunnel barrier due to larger current density complicates the interpretation of these measurements. Furthermore, our results cannot be accounted for by the difference in temperature of the free layer in the P and AP states that results from differences in the thermal conductivity of the MTJ in these two states, since this leads to changes in temperature that are much too small to account for our observations (*27*). We note that this would mean, for example, that the free layer of the MTJ in the AP configuration would have to be 15 K hotter (dashed black line in Fig. 2e) than in the P configuration, for $I_H = 0.15 mA$, which is unrealistic. We can also obtain the $H_{sw}$ from the magneto-Seebeck measurements for $|I_H| \geq 0.1 mA$, where the P → AP and AP → P switching can be ascertained (Fig. 4b). Even though the thermoelectric charge current vanishes in the open circuit configuration, we find evidence of the thermospin current (*9*) that exerts TST thereby influencing $H_{sw}^{AP \to P}$ but not $H_{sw}^{P \to AP}$ (pink symbols in Fig. 2e,f) for device I. This is why the $H_{sw}^{AP \to P}$ and $H_{sw}^{P \to AP}$ in the open-circuit configuration are the same, within experimental error, to those measured in the closed-circuit configuration for the same temperature gradient.

In order to ascertain the origin of the TST, we perform similar experiments on another device III (Fig. 5), fabricated with a different magnetic stack grown on MgO (100) substrate with the layer sequence from bottom to top as: 7.5 Ta | 12.5 IrMn | 0.6 CoFeB | 3 CoFe | 0.4 Ru | 2.7 CoFe | 0.9 MgO | 2 CoFeB | 5 Ta | 5 Ru, where the numbers represent film thicknesses in nanometers. Here the magnetic electrodes of the MTJ adjacent to the MgO tunnel barrier are different from devices I and II. Performing the same exercise (Fig. 5c,d) of comparing $H_{sw}(T+\Delta T)$ in both open and closed circuit configurations with $H_{sw}(T)$, we see that all the measurements for device III lie on top of one another, once again confirming our temperature estimation procedure and also showing no appreciable TST for either AP → P or P → AP switching, even though the TMR (~ 126 %) for device III (Fig. 5b) is almost five times higher than the previous devices (the RA of device III is ~ 6 Ω-μm², same as devices I and II). We show that the TST depends on the current-voltage (IV) characteristics of the MTJ instead. Fig. 5e shows the normalized tunneling conductance $G_{norm}(V) = \frac{G(V)}{G(V \to 0)}$, where $G(V) = \frac{I(V)}{V}$ for devices II and III in their respective AP and P states. For device II (and I) since there is an asymmetry in the $G_{norm}$ across $V=0$ in the AP state, we observe the evidence of the TST affecting AP → P switching, whereas negligible TST is found for P → AP switching in device II (and I), and both AP → P and P → AP switching of device III, as the $G_{norm}$ is much more symmetric across $V=0$ in these cases (Fig. 5f). This asymmetry indicates a change in the tunnel spin polarization (*4, 18*) of the tunnel barrier as a function of energy near zero bias voltage, which could stem from, for example, changes in the tunneling matrix elements or variations in the local density of states of the magnetic electrodes at the tunnel interface. Furthermore, we show that the asymmetry in $G_{norm}$ is consistent with minority spins from the reference layer accumulating in the free layer, when the free layer is hotter than the reference layer in the AP configuration as is the case in our experiments (see Supplementary Information SVI).

In summary, we have shown that temperature gradients of ~1 K/nm across an ultra-thin tunnel barrier can induce large spin currents and thereby a giant TST that can influence MTJ switching. The measurements reported here are performed with static temperature gradients. Much sharper temperature gradients can be created on short time scales to create greater TST, which might be large enough to switch an MTJ with pure temperature gradients alone thereby making it relevant to the Magnetic Random Access Memory (MRAM) technology (*28*). We postulate that the TST can be enhanced even further by the appropriate choice of asymmetric ferromagnetic electrodes, such as Heusler alloys, so as to enhance the bias voltage dependence of the tunnel spin polarization.

**Figure Captions**

**Fig. 1. Device geometry. a**, (Top) Scanning Electron Micrograph (SEM) of the device I showing the free and reference layers of the MTJ along with the gold contacts and the heater in the blow-up in the red box. (Bottom) Cross-section Transmission Electron Micrograph (X-TEM) showing from bottom to top: MgO (100) substrate, the vertical magnetic stack, the 30 nm thick top Au contact to the free layer, the 20 nm thick AlOx isolation pad and the 20 nm thick ScN heater. Blow-up in green box shows the magnetic stack. **b**, Schematic showing the various components of the device structure, the electrical measurement circuit in the closed-circuit configuration, and the heat flow direction. **c**, COMSOL model built from the SEM and TEM information of the device showing in **d**,

the temperature profile for different $I_H$s along the z-axis of the entire stack centered on the MTJ (white dashed arrow in c). The inset shows the temperature profile across the magnetic electrodes and the tunnel barrier. **e**, The temperature profile of the bottom reference layer (z = 142 nm) along the x-axis used for thermometry (yellow dashed line in c).

**Fig. 2. MTJ switching measurements and local thermometry for device I. a**, (Inset) TMR of device I at 10 K with the $H_{sw}$ indicated. $H_{sw}^{AP \to P}$ and $H_{sw}^{P \to AP}$ plotted as a function of $I_H$ in a and **b**, respectively. **c & d**, Reference layer resistance change as a function of heater current $\Delta R_{RL}(I_H)$ and temperature as a function of free layer resistance change $T(\Delta R_{RL})$ are used to estimate $T(I_H)$ with an appropriate scaling factor (see Supplementary Information SIII). $\Delta T(I_H)$ is obtained from the COMSOL model. $T(I_H)$ is then used to plot $H_{sw}(T + \Delta T)$ along with independently measured $H_{sw}(T)$ for AP → P and P → AP switching in **e & f**. Pink lines indicate $H_{sw}(T + \Delta T)$ in an open circuit (O.C.) configuration discussed further in Fig. 4.

**Fig. 3. MTJ switching measurements for device II. a**, SEM of device II (same dimensions as device I) with the free layer oriented at 120° to the reference layer. **b**, TMR of device II along with the different magnetization orientations (A through G) of the free layer with respect to the reference layer. **c**, $H_{sw}(T + \Delta T)$ and $H_{sw}(T)$ for orientations A, B and G, where evidence of $\Delta T(I_H)$ induced spin currents effects are observed. $H_{sw}(T + \Delta T)$ and $H_{sw}(T)$ for orientations C, D, E and F, where $H_{sw}(T + \Delta T)$

and $H_{sw}(T)$ track each other implying no $\Delta T(I_H)$ induced spin currents effects, while also confirming our temperature estimation.

**Fig. 4. Magneto-Seebeck measurements of devices I and II. a**, Schematic showing the open-circuit measurement configuration (i.e. when $I_{MTJ} = 0$). **b & c**, Seebeck voltage, $V_S$, as a function of $H$ for $I_H = +0.2mA$ for device I and II respectively. **d & e**, $V_S$ as a function of $I_H$ in the AP and P configuration of the device I and II respectively. The tunneling magnetothermopower (TMTP) ratio is defined as $\frac{V_S^{AP} - V_S^P}{V_S^{AP}}$.

**Fig. 5. Origin of the TST. a**, TEM of device III showing the 1 nm thick MgO tunnel barrier. The free layer of the MTJ is 185 nm long and 65 nm wide, whereas the reference layer is of the same dimensions as devices I and II. **b,** TMR of device III. $H_{sw}$ are larger for device III because of smaller dimensions of the free layer and TMR is higher because of different magnetic electrodes. **c & d**, $H_{sw}(T + \Delta T)$ along with independently measured $H_{sw}(T)$ for AP → P and P → AP switching for device III. Pink lines indicate $H_{sw}(T + \Delta T)$ in an open circuit (O.C.) configuration as discussed in Fig. 4. **e**, $G_{norm}$ for devices II and III in AP and P states. Data near $V = 0$ has been taken out. The table shows that TST is present whenever there is a strong asymmetry in $G_{norm}$ across $V = 0$.

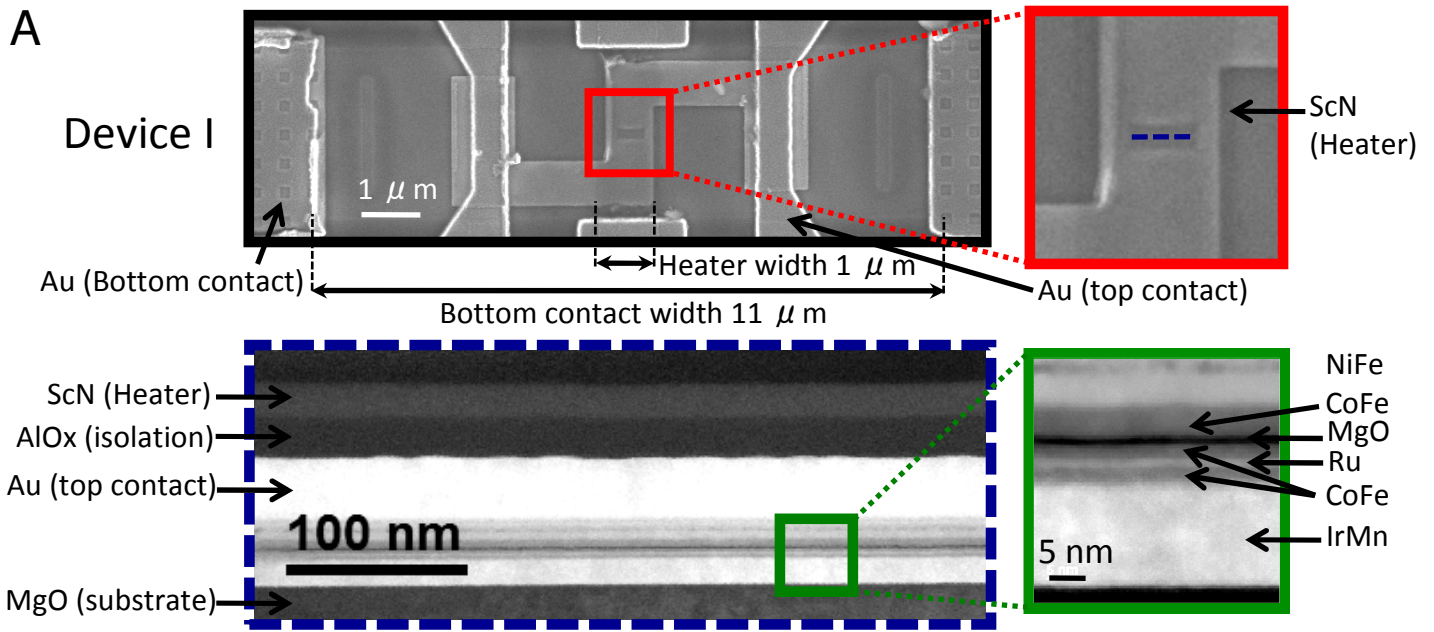
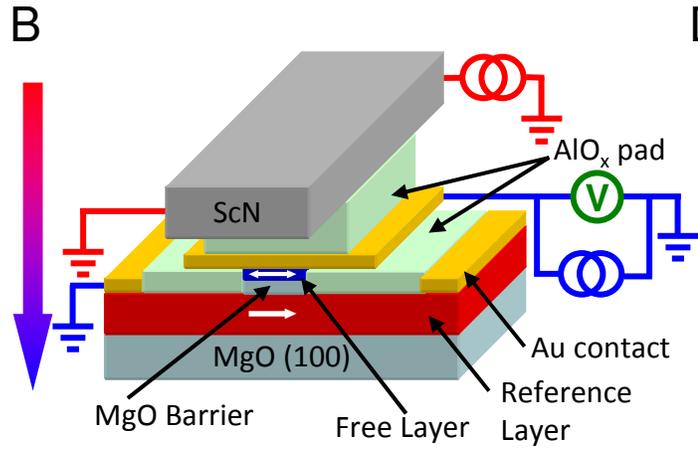
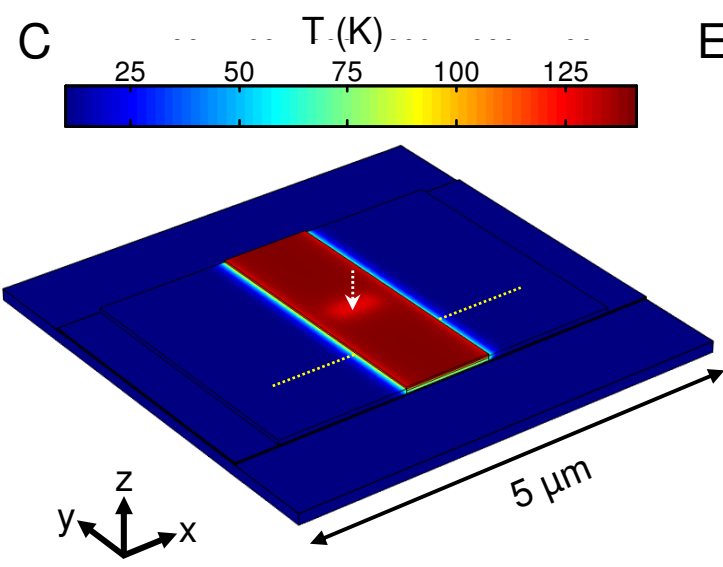
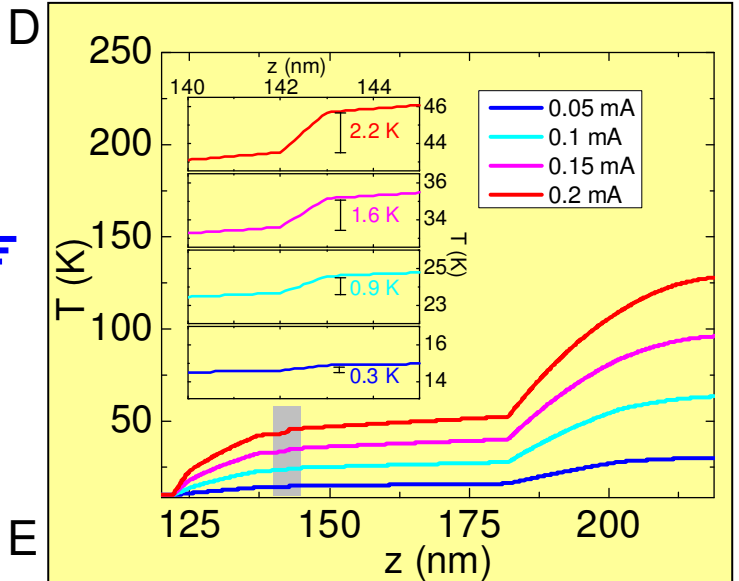
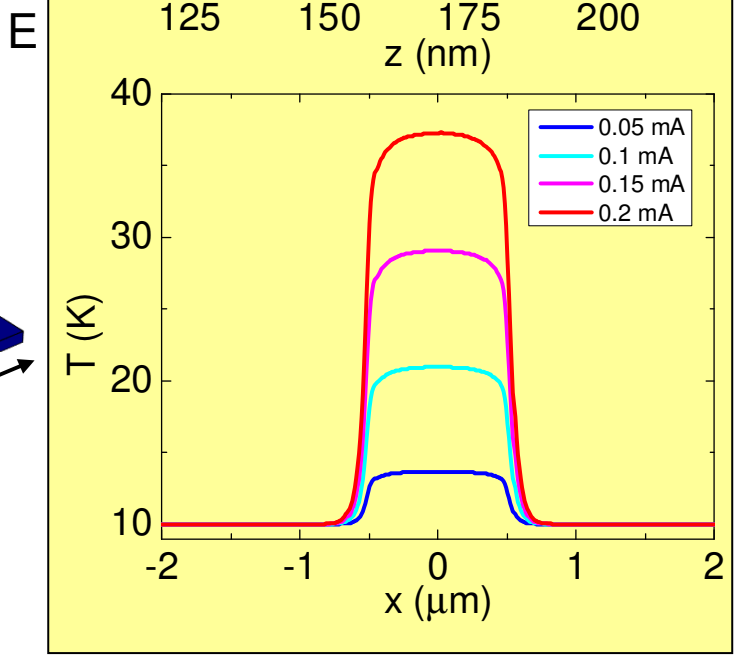

Figure 1 Pushp, Phung et al.

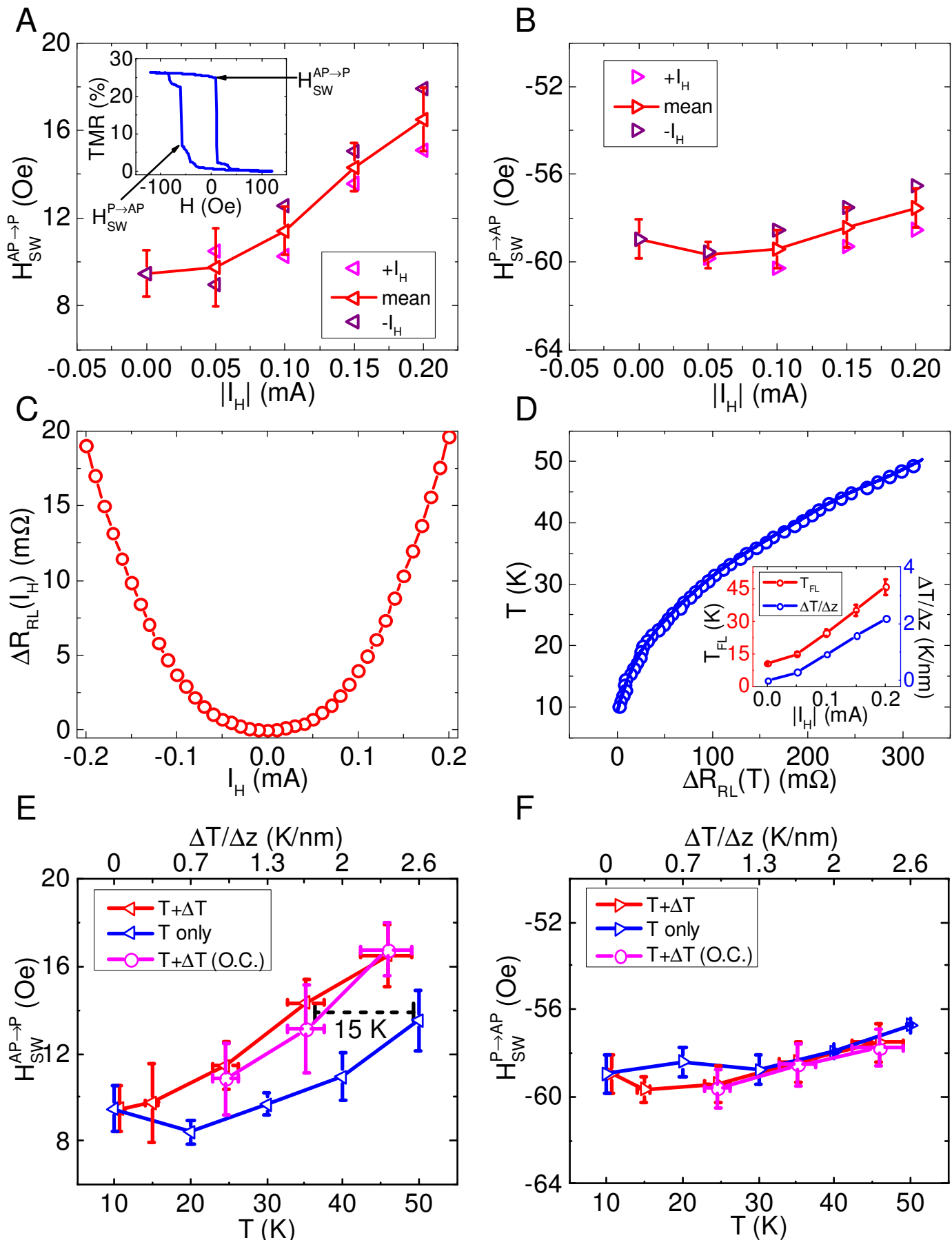

Figure 2 Pushp, Phung et al.

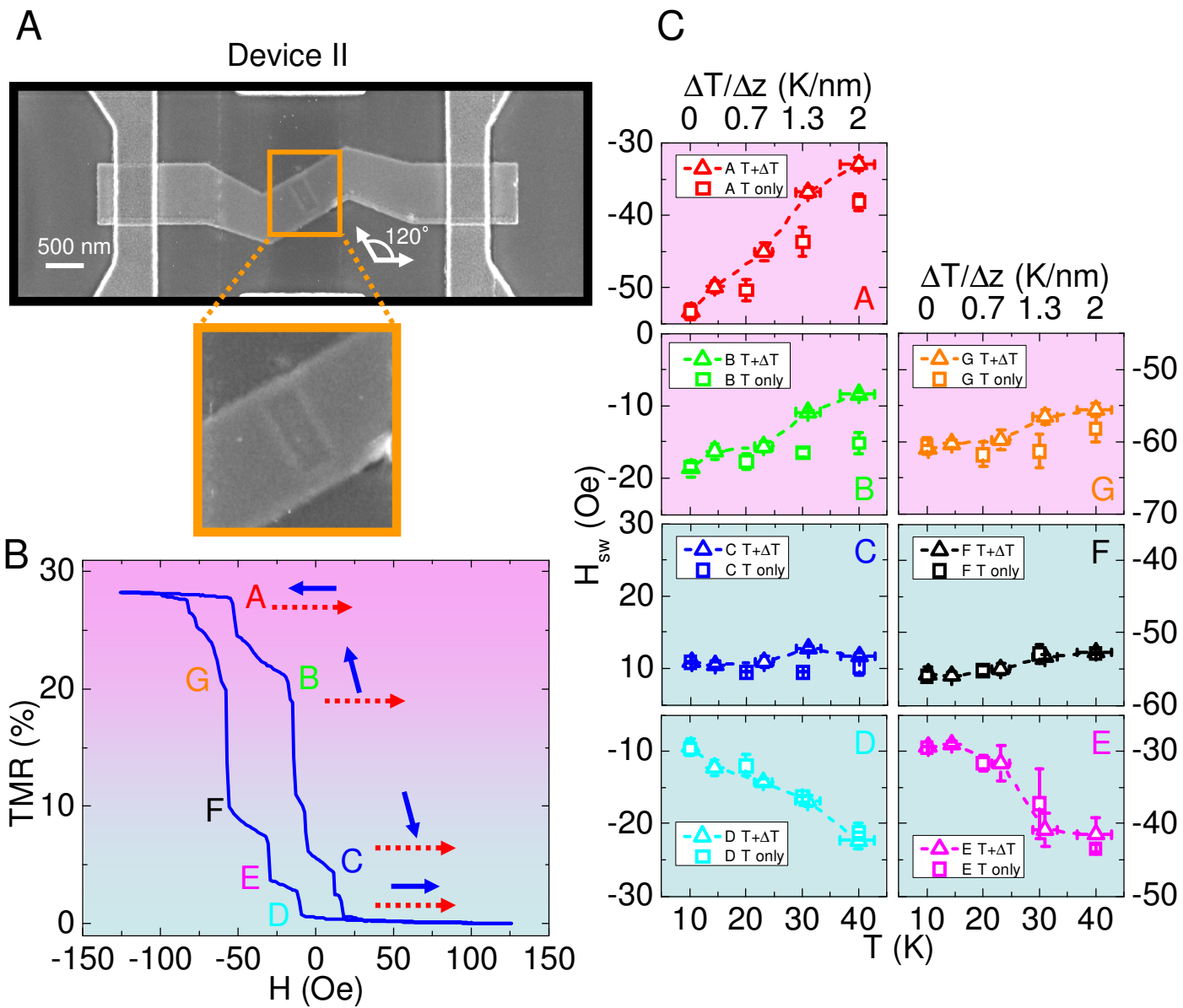

Figure 3 Pushp, Phung et al.

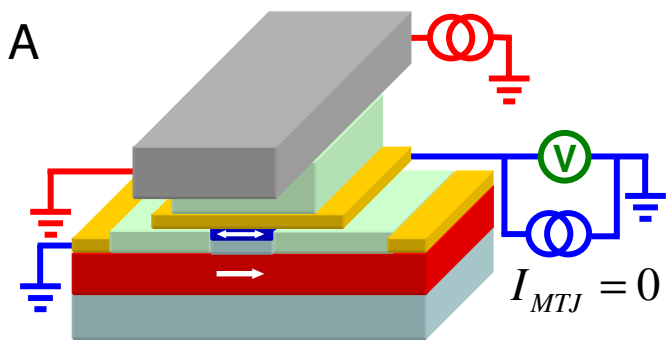

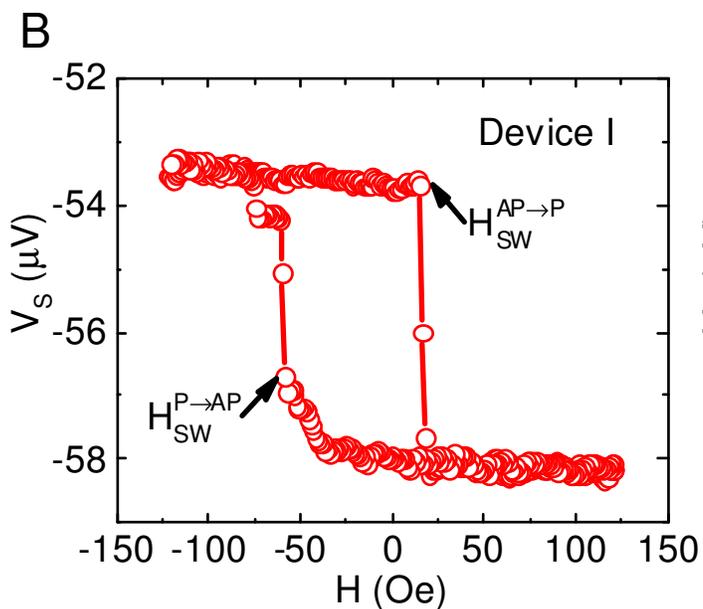
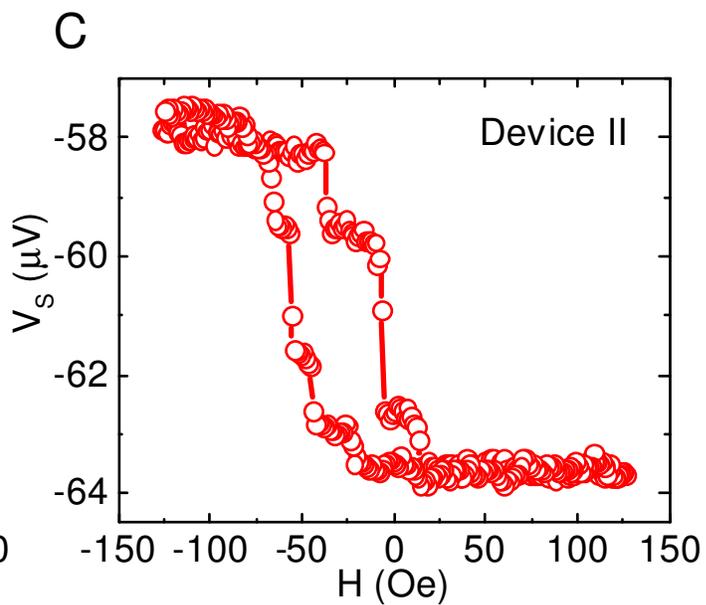
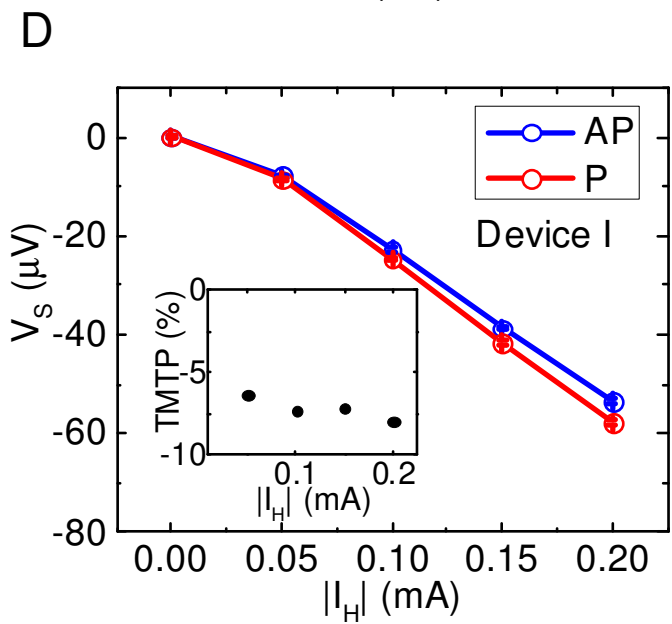
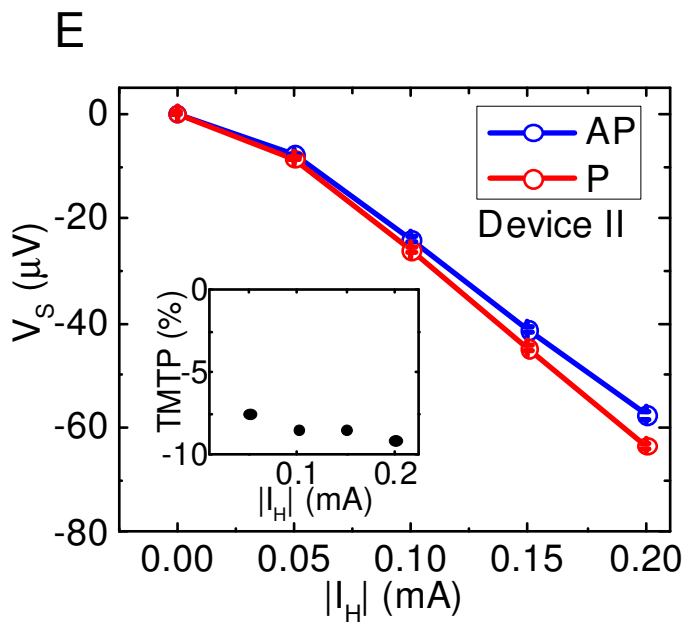

Figure 4 Pushp, Phung et al.

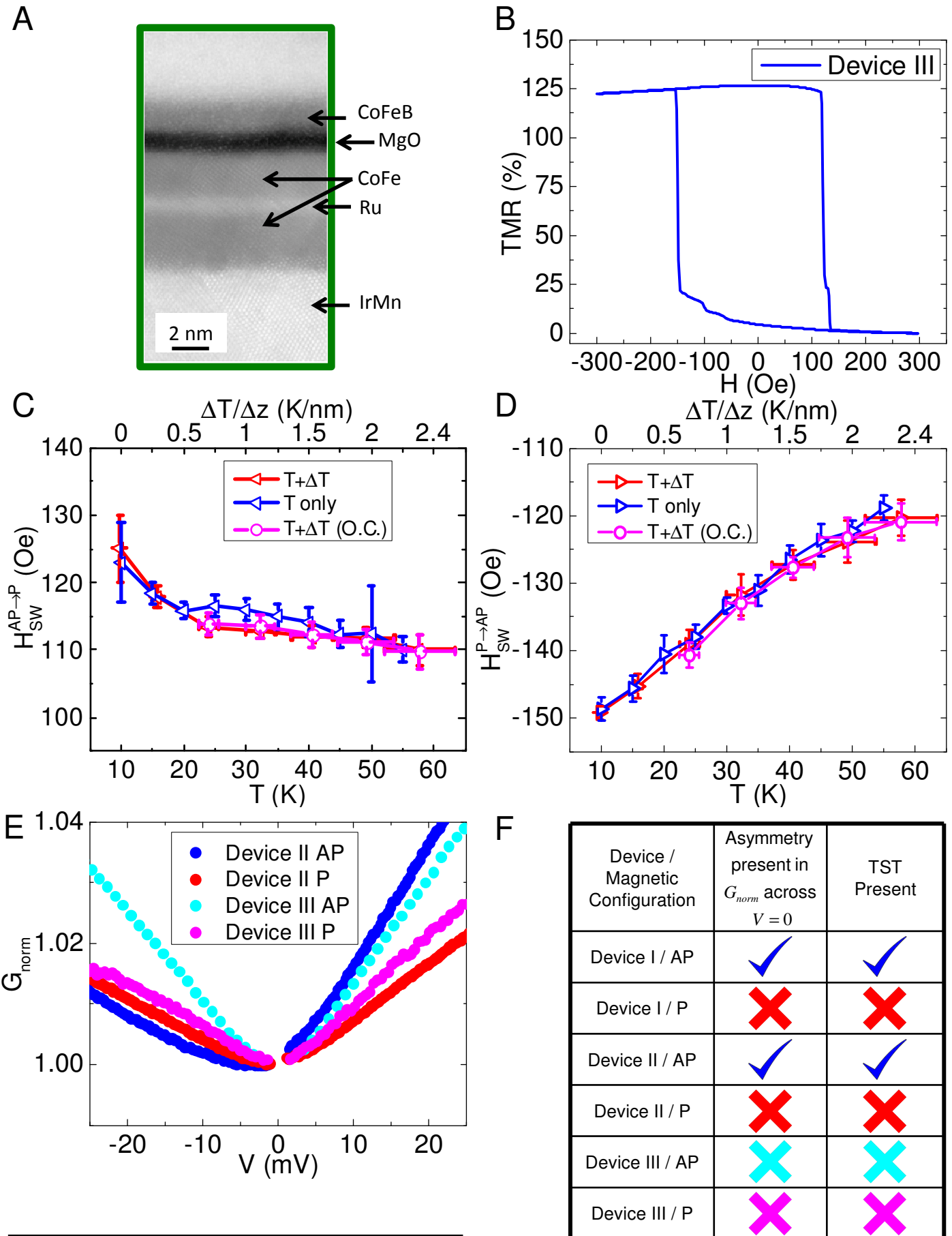

Figure 5 Pushp, Phung et al.